# Deformation of the J-matrix method of scattering

A. D. Alhaidari

*Physics Department, King Fahd University of Petroleum & Minerals, Box 5047, Dhahran 31261, Saudi Arabia*
E-mail: **haidari@mailaps.org**

We construct nonrelativistic J-matrix theory of scattering for a system whose reference Hamiltonian is enhanced by one-parameter linear deformation to account for nontrivial physical effects that could be modeled by a singular ground state coupling.

In their efforts to explain scattering data by local interaction, theoretical physicists are sometimes faced with the prospect of trying to account for a highly nontrivial physical effect, which is so prominent that it transcends all potential models that they could come up with. Very often in such cases, a background interaction in the reference Hamiltonian $H_0$ could be introduced that may bring about some improvements on the theoretical results [1]. However, rarely one could find an analytic form of the interaction model that could account for this phenomenon to ones satisfaction. Most importantly, even if this is done and found to be partially successful, the full effect of such background interaction, which might have induced substantial improvements on the results, is usually not included. This is due to the fact that in most of these theoretical calculations, the background interaction is confined to a finite region in configuration or function space and not taken in totality when obtaining the full solution to the reference problem [2]. The complication lies in trying to find an analytic solution of the combined reference problem that may prove to be very difficult to handle. Our contribution in this letter is to improve on this last point for a particular class of interactions. We are proposing to construct a deformation of the J-matrix theory of scattering [3] in which we could take into account the total effect of this deformation in the full analytic solution of the $H_0$-problem. We consider a one-parameter linear deformation that is suitable for the interpretation of physical effects that could be modeled by singular ground state coupling. However, due to our lack of know-how in the field of data fitting and limited expertise to mathematical physics, no specific scattering data will be the target of our investigation. Instead, we will only be concerned with the development of the formalism, which will then be applied to a potential scattering example where the effect of such deformation is demonstrated. This effect will be calculated by taking the full contribution of the deformation in the reference Hamiltonian and compared to that where the effect is limited to a finite region in function space to which the short-range model potential is confined. We start by a brief qualitative introduction to the essentials of the J-matrix theory of scattering, which are necessary for this development.

The J-matrix is an algebraic method of quantum scattering with substantial success in atomic and nuclear physics [4]. The accuracy and convergence property of the method compares favorably with other successful scattering calculation methods [5]. Its structure in function space is endowed with formal and computational analogy to the R-matrix method [6] in configuration space. The $L^2$ basis in the J-matrix method is chosen such that it can support a tridiagonal matrix representation for the reference

Hamiltonian and the identity. This property allows for a full algebraic solution to the $H_0$-problem. It is obtained as an analytic solution of the symmetric three-term recursion relation resulting from the tridiagonal structure of the matrix representation of the reference wave equation. The J-matrix method yields exact scattering information over a continuous range of energy for a model potential obtained by truncating the given short-range potential in a finite subset of this basis. It was shown to be free from fictitious resonances that plague some algebraic variational scattering methods [7]. It has also been extended to multi-channel [8] as well as relativistic [9] scattering. Recently, a modification to the formalism was introduced to account for coupling to long-range potentials [10].

The J-matrix aims at giving an algebraic solution to the problem whose time independent Schrödinger wave equation is
$$\left(H_0 + \tilde{V} - E\right)|\chi(E)\rangle = 0$$
where the wave function $\chi$ is energy normalized, that is $\langle \chi(E)|\chi(E')\rangle = \delta(E - E')$, and belongs to an $L^2$ space whose basis elements, $\{\phi_n\}_{n=0}^{\infty}$, are chosen such that they produce a tridiagonal matrix representation for the operator $J(E) \equiv H_0 - E$. The short-range model potential $\tilde{V}$ is assumed to be well represented by its matrix elements in a finite subset of this basis $\{\phi_n\}_{n=0}^{N-1}$ for some large enough integer $N$. On the other hand, the reference problem whose wave equation is written as $(H_0 - E)|\psi(E)\rangle = 0$, where $|\psi(E)\rangle = \sum_n d_n(E)|\phi_n\rangle$, has a full analytic solution. That is the symmetric three-term recursion relation resulting from the equivalent matrix wave equation

$$\sum_{m=0}^{\infty} J_{nm}(E) d_m(E) = 0 \; ; \quad n \geq 0 \qquad (1)$$

is solved analytically for the expansion coefficients $\{d_n(E)\}_{n=0}^{\infty}$ of the reference wave function. Typically, there are two sets of such solutions to the $H_0$-problem; one is regular at the origin while the other is not. Regularization gives two wave functions that behave asymptotically as sinusoidal with a phase difference of $\pi/2$. The corresponding expansion coefficients are $d_n(E) = s_n(E)$ for the regular sine-like solution and $d_n(E) = c_n(E)$ for the regularized cosine-like solution. The analytic expressions for these coefficients in the J-matrix bases that are suitable for several reference problems are found and tabulated elsewhere in the literature [11]. The initial relations imposed by regularization on the recursion (1) are

$$\begin{aligned} J_{00} s_0 + J_{01} s_1 &= 0 \\ J_{00} c_0 + J_{01} c_1 &= k/2 s_0 \end{aligned} \qquad (2)$$

where $k = \sqrt{2E}$. The remaining relations are
$$J_{n,n-1} d_{n-1} + J_{n,n} d_n + J_{n,n+1} d_{n+1} = 0 \; ; \quad n \geq 1 \qquad (3)$$
where $d_n$ stands for either $s_n$ or $c_n$. We choose to work instead with the complex coefficients defined as $h_n^{\pm}(E) \equiv c_n(E) \pm i s_n(E)$ and with the J-matrix kinematical factors obtained from these as the ratios $T_n(E) = h_n^-(E)/h_n^+(E)$ and $R_{n+1}^{\pm}(E) = h_{n+1}^{\pm}(E)/h_n^{\pm}(E)$. The homogeneous three-term recursion (3) and its initial relations (2) can now be rewritten in terms of these complex coefficients as

$$J_{00}h_0^\pm + J_{01}h_1^\pm = ik/(h_0^\pm - h_0^\mp) \tag{2'}$$

$$J_{n,n-1}h_{n-1}^\pm + J_{n,n}h_n^\pm + J_{n,n+1}h_{n+1}^\pm = 0 \;;\quad n \geq 1 \tag{3'}$$

Now a linear deformation in the recursion, whose physical implications to the $H_0$-problem will be interpreted shortly, is introduced as follows:

$$J_{00}\hat{h}_0^\pm + J_{01}\hat{h}_1^\pm = -\mu\hat{h}_0^\pm + ik/(\hat{h}_0^\pm - \hat{h}_0^\mp) \tag{4}$$

where $\mu$ is the real deformation parameter whose dimension is that of energy. All quantities with the caret symbol refer to the new problem, which are obtained from the original ones by this deformation. The homogenous recursion relation (3') is invariant under this deformation (i.e. it remains the same in terms of the coefficients $\hat{h}_n^\pm$). To obtain the full analytic solution to the $\hat{H}_0$-problem, we propose the following transformation of the coefficients

$$\hat{h}_n^\pm(\mu,E) = e^{\pm i\tau(\mu,E)} h_n^\pm(E) \tag{5}$$

such that the original $H_0$-problem is recovered. $\tau(\mu,E)$ is a real energy-dependent phase angle parameterized by the deformation constant $\mu$. This transformation induces the following change in the J-matrix kinematical factors

$$\hat{T}_n = e^{-2i\tau}T_n \;;\quad \hat{R}_{n+1}^\pm = R_{n+1}^\pm \qquad;n\geq 0$$

Substituting from (5) into (4) and using (2') we obtain the following analytic expression for the transformation phase

$$e^{2i\tau} = T_0 + (1-T_0)\left[1 + \frac{\mu}{J_{00} + J_{01}R_1^+}\right]^{-1} \tag{6}$$

It is evident that the linear deformation introduced in (4) is equivalent to the transformation in the reference Hamiltonian $H_0 \to \hat{H}_0$, where

$$\left(\hat{H}_0\right)_{nm} = \left(H_0\right)_{nm} + \mu\delta_{n0}\delta_{m0} \tag{7}$$

This could be interpreted as a singular effect coupled only to the ground state. Following the standard J-matrix scattering calculations [3], we arrive at the following expression for the $N^{th}$ order scattering matrix:

$$\hat{S}^{(N)}(\mu,E) = e^{-2i\tau(\mu,E)} T_{N-1}(E) \frac{1 + \hat{g}_{N-1,N-1}(\mu,E) J_{N-1,N}(E) R_N^-(E)}{1 + \hat{g}_{N-1,N-1}(\mu,E) J_{N-1,N}(E) R_N^+(E)} \tag{8}$$

where the finite Green's function

$$\hat{g}_{N-1,N-1}(\mu,E) = \left\langle \overline{\phi}_{N-1} \left| \left(\hat{H}_0 + \tilde{V} - E\right)^{-1} \right| \overline{\phi}_{N-1} \right\rangle$$

is evaluated in the $L^2$ conjugate space whose elements $\{\overline{\phi}_n\}_{n=0}^\infty$ satisfy $\langle \overline{\phi}_n | \phi_m \rangle = \delta_{nm}$.

Now, we apply this deformation to a problem whose reference Hamiltonian includes the Coulomb interaction $Z/r$ and consider single channel scattering by the potential $\tilde{V}(r) = 7.5 r^2 e^{-r}$ which is known to have an S-wave Coulomb-free sharp resonance at $E_r = 3.426$ in atomic units [12]. The results of scattering calculation using this J-matrix deformation formalism are shown graphically as plots of $|1-\hat{S}|$ versus the energy $E$. The solid curve is obtained by taking the full effect of the deformation into account as given by the scattering matrix in equation (8). The dashed curve, on the other hand, is obtained by considering only the truncated $\hat{H}_0$-problem which is equivalent to the $H_0$-problem except that $\tilde{V}_{nm} \to \tilde{V}_{nm} + \mu\delta_{n0}\delta_{m0}$, where $n,m = 0,1,\ldots,N–1$. That is, the scattering matrix for the truncated problem is that given by the expression in (8) except that the phase factor $e^{-2i\tau}$ is missing. The calculation was done in the J-matrix Laguerre basis [3] with a basis scale parameter $\lambda = 5.0$ while the physical parameters were taken as $Z = 0$, $l = 0$, and $\mu = 1.0$. Figure 1(a) gives the results for a subspace whose dimension $N = 20$. It is evident that truncating the effect of deformation fails to reproduce, or approach, the right results even if $N$ were to be increased to computationally manageable dimensions as signified by the plots in Figure 1(b) and 1(c) where $N = 30$ and 50, respectively. However, prominent physical effects (for example, the sharp resonance at $E_r = 3.62$ a.u., which was clearly shifted by the deformation) are captured in both calculations. Figure (2) is a plot of the transformation phase angle $\tau(\mu,E)$ given by formula (6), in radians, as a function of energy for the parameters whose values are stated above.

Finally, it is worthwhile to note that higher order linear deformation is also possible. We are currently investigating this problem where three- and six-parameter deformation has been achieved and results are to be published shortly elsewhere. Here, it might be appropriate and sufficient to highlight some of the main results obtained for the three-parameter deformation. In this case and in analogy with Eq. (7) for the one-parameter deformation, the reference Hamiltonian transforms as follows:

$$\hat{H}_0 = H_0 + \begin{pmatrix} \mu_+ & \mu_0 \\ \mu_0 & \mu_- \end{pmatrix}$$

which could be interpreted as an effect originating from close coupling of the lowest states. The resulting transformation phase needed for the evaluation of the scattering matrix in equation (8) is

$$e^{2i\tau} = \left(\frac{J_{01} + \mu_0}{J_{01} - \mu_- R_1^+}\right)^2 \left\{ T_0 \left|\frac{J_{01} - \mu_- R_1^+}{J_{01} + \mu_0}\right|^2 + (1-T_0) \times \right.$$
$$\left. \left(J_{00} + J_{01}R_1^+\right)\left[J_{00} + \mu_+ + R_1^+ \frac{(J_{01} + \mu_0)^2}{J_{01} - \mu_- R_1^+}\right]^{-1} \right\}$$

Furthermore, an *exact* numerical solution was also obtained for the problem with three-parameter deformation defined by

$$\left(\hat{H}_0\right)_{nm} = \left(H_0\right)_{nm} + \mu_+\delta_{n0}\delta_{m0} + \mu_-\delta_{n,M}\delta_{m,M} + \mu_0\delta_{n0}\delta_{mM} + \mu_0\delta_{nM}\delta_{m0} \qquad (9)$$

for a given positive integer $M$. This could also be interpreted as singular coupling of two non-neighboring states (e.g., in a model for a crystal with coupling between two separated lattice sites). Figure (3) gives the resulting transformation phase angle in this case with the following parameters $Z = 1$, $l = 0$, $M = 7$, $\lambda = 5.0$, and

$$\mu_+ = 1.0, \mu_- = 0.5, \mu_0 = -0.7 \qquad (10)$$

## ACKNOWLEDGEMENT

The author is indebted to Dr. H. A. Yamani for insight into the physical interpretation of the deformation introduced in this work.

**REFERENCES**

[1] This could be inferred by surveying the physics literature especially in nuclear and condensed matter. See, for example, A. D. Barut (editor), "Scattering theory; new methods and problems in atom, nuclear and particle physics", (Gordon and Breach, London, UK, 1969); D. E. Pritchard, J. Chem. Phys. **56**, 4206 (1972); B. Buck, H. Friedrich and C. Wheatley, Nucl. Phys. A **275**, 246 (1977); L. J. Allen, Phys. Rev. A **34**, 2706 (1986); V. I. Kukulin, V. N. Pomerantsev and J. Horacek, Phys. Rev. A **42**, 2719 (1990); H. F. Arellano, F. A. Brieva, W. G. Love and K. Nakayama, Phys. Rev. C **43**, 1875 (1991); V. Stoks and J. J. de Swart, Phys. Rev. C **52**, 1698 (1995).

[2] Examples are found in systems confined to a box in configuration space or to finite $L^2$ bases such as the Fredholm method [see, for example, T. S. Murtaugh and W. P. Reinhardt, Chem. Phys. Lett. **11**, 562 (1971)]

[3] E. J. Heller and H. A. Yamani, Phys. Rev. A **9**, 1201 (1974); **9**, 1209 (1974); H. A. Yamani and L. Fishman, J. Math. Phys. **16**, 410 (1975); J. T. Broad and W. P. Reinhardt, Phys. Rev. A **14**, 2159 (1976).

[4] See, for example, G. F. Filippov and I. P. Okhrimenko, Yad. Fiz. **32**, 932 (1980); **33**, 928 (1981) [Sov. J. Nucl. Phys. **32**, 480 (1980); **33**, 488 (1981)]; Yu. F. Smirnov and Yu. I. Nechaev, Kinam **4**, 445 (1982); Yu. I. Nechaev and Yu. F. Smirnov, Yad. Fiz. **35**, 1385 (1982) [Sov. J. Nucl. Phys. **35**, 808 (1982)]; J. Revai, M. Sotona, and J. Zofka, J. Phys. G **11**, 745 (1985).

[5] See, for example, W. P. Reinhardt, Comp. Phys. Comm. **17**, 1 (1979); J. T. Broad, Phys. Rev. A **31**, 1494 (1983); I. Bray and A. T. Stelbovics, Phys. Rev. Lett. **69**, 53 (1992); D. A. Konovalov and I. E. McCarthy, J. Phys. B **27**, L407 (1994); I. Bray, I. E. McCarthy, and A. T. Stelbovics, J. Phys. B **29**, L245 (1996); W. H. Kuan, T. F. Jiang, and K. T. Chung, Phys. Rev. A **60**, 364 (1999).

[6] See, for example, A. M. Lane and R. G. Thomas, Rev. Mod. Phys. **30**, 257 (1958); A. M. Lane and D. Robson, Phys. Rev. **178**, 1715 (1969).

[7] E. J. Heller, Phys. Rev. A **12**, 1222 (1975).

[8] J. T. Broad and W. P. Reinhardt, J. Phys. B **9**, 1491 (1976); H. A. Yamani and M. S. Abdelmonem, J. Phys. B **30**, 1633 (1997); **30**, 3743 (1997).

[9] P. Horodecki, Phys. Rev. A **62**, 052716 (2000); A. D. Alhaidari, H. A. Yamani and M. S. Abdelmonem, Phys. Rev. A **63**, 062708 (2001); A. D. Alhaidari, J. Math. Phys. **43**, 1129 (2002).

[10] W. Vanroose, J. Broeckhove, and F. Arickx, Phys. Rev. Lett. **88**, 010404 (2001).

[11] J. T. Broad, Phys. Rev. A **18**, 1012 (1978); P. C. Ojha, J. Phys. A **21**, 875 (1988).

[12] V. A. Mandelshtam, T. R. Ravuri and H. S. Taylor, Phys. Rev. Lett. **70**, 1932 (1993); H. A. Yamani and M. S. Abdelmonem, J. Phys. A **28**, 2709 (1995).

**FIGURE CAPTIONS**

FIG. 1: $\left|1-\hat{S}(\mu,E)\right|$ vs. the energy $E$ for scattering by the short range potential $\tilde{V}(r)=7.5r^2e^{-r}$ with the deformed reference Hamiltonian given by Eq. (7). The dashed curve is the result of calculation for a truncated $\hat{H}_0$-problem, while the solid curve is obtained by taking the full effect into account. The subspace dimension in Figure (a), (b), and (c) is taken as $N=20$, 30, and 50, respectively. The physical parameters were set to $Z=0$, $l=0$, and $\mu=1.0$ a.u. while the Laguerre basis scale parameter $\lambda=5.0$.

FIG. 2: The transformation phase angle $\tau(\mu,E)$ in radians as a function of energy for the problem whose physical parameters are listed in the caption of Fig. 1.

FIG. 3: The result of numerical evaluation of the transformation phase angle as a function of energy for the three-parameter deformation defined by Eqs. (9) and (10). The remaining parameters used in the calculation are $Z=1$, $l=0$, $M=7$, and $\lambda=5.0$.

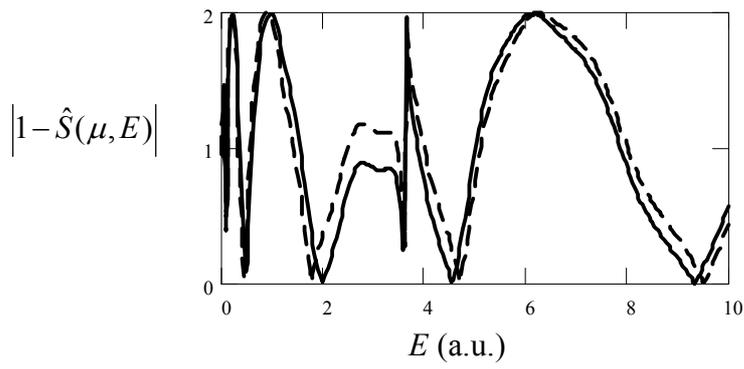

Fig. 1(a)

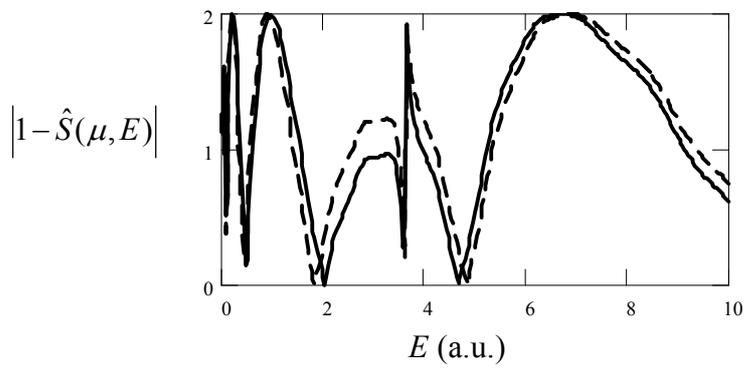

Fig. 1(b)

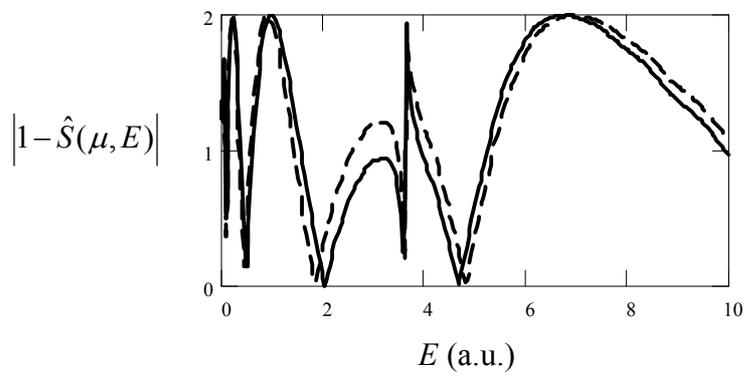

Fig. 1(c)

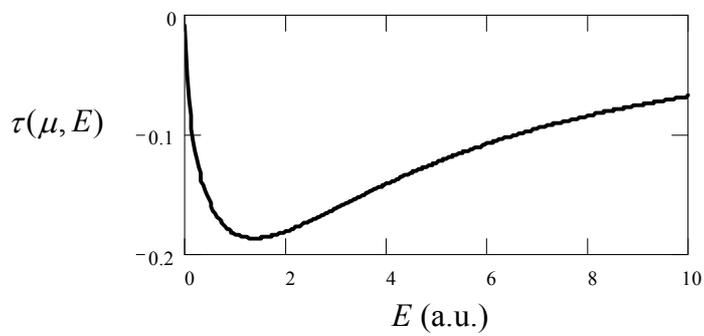

Fig. 2

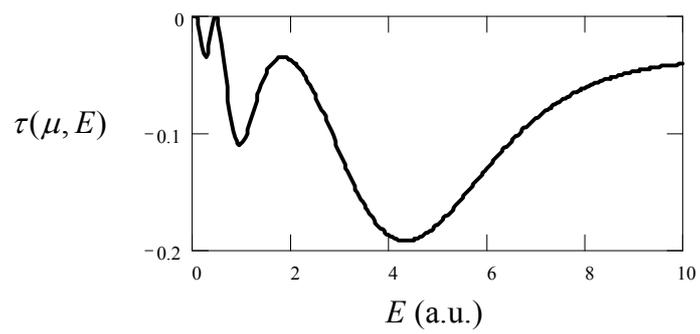

Fig. 3